\pdfoutput=1
\documentclass[%
reprint,
superscriptaddress,
 amsmath,amssymb,
 aps,
]{revtex4-2}

\usepackage{graphicx}
\usepackage{dcolumn}
\usepackage{bm}
\usepackage{here}
\usepackage{float}
\usepackage{physics}

\begin{document}

\preprint{APS/123-QED}

\title{Tensor network renormalization approach to antiferromagnetic 6-state clock model on the Union Jack lattice}

\author{Kenji Homma}
\affiliation{%
 Institute for Solid State Physics, The University of Tokyo, Kashiwa, Chiba 277-8581, Japan
}%

\author{Satoshi Morita}

\affiliation{%
Graduate School of Science and Technology, Keio University, Yokohama, Kanagawa 223-8522, Japan \looseness=-1
}%
\author{Naoki Kawashima}%
\affiliation{%
 Institute for Solid State Physics, The University of Tokyo, Kashiwa, Chiba 277-8581, Japan
}%
\affiliation{%
Trans-scale Quantum Science Institute, The University of Tokyo, Bunkyo, Tokyo 113-0033, Japan
}%
\date{\today}
\begin{abstract}
Using the nuclear norm regularization techniques on tensor network renormalization algorithm, we study the phase diagram, the critical behavior and the duality property of 
the antiferromagnetic 6-state clock model on the Union Jack lattice. We find that this model undergoes multiple phase transitions; there is the Berezinskii-Kosterlitz-Thouless, 
$Z_{6}$ symmetry breaking and chiral transition with decreasing temperature. Furthermore, we provide convincing numerical evidence that its quasi-long range order is well explained 
by the compactified boson conformal field theory (CFT) and the chiral transition is in perfect agreement with the Ising CFT, including central charge, scaling dimension spectrum, 
and operator product expansion coefficients.

\end{abstract}
\maketitle
\section{Introduction}
The critical behavior of frustrated spin systems has been investigated quite intensively, due to the exotic critical behavior and phase diagrams. 
Among these frustrated spin systems, many numerical studies have been performed to elucidate the criticality of fully frustrated XY models 
\cite{doi:10.1143/JPSJ.53.1145,PhysRevB.42.2438,PhysRevB.49.9567,PhysRevB.49.15184,PhysRevLett.75.2758,PhysRevB.58.5163,PhysRevB.50.1061}. 
The typical example is the antiferromagnetic (AF) XY model on the triangular lattice. Apart from the global $U(1)$ internal symmetry, 
this model is known to exhibit a $Z_{2}$ symmetry breaking transition originating from macroscopically degenerated ground states induced by 
geometrical frustration. Despite the controversy over the nature of its critical properties, it is generally accepted that the phase transition associated 
with the quasi-long range order (QLRO), as known as the Berezinskii-Kosterlitz-Thouless (BKT) transition, lies considerably close to, but strictly lower than, 
the Ising-like transition with $Z_{2}$ symmetry breaking, namely $T_{\text{BKT}}<T_{\text{Is}}$.

Although all classical AFXY models on the lattice consisting of the triangular plaquette can be categorized as the fully frustrated XY models, their critical properties 
may depend on the lattice structure. For example, the AFXY models on the Union Jack (UJ) lattice are known to have the reverse order of transitions as 
$T_{\text{BKT}}>T_{\text{Is}}$ \cite{PhysRevB.87.024108,PhysRevB.96.144432}. It is expected that the critical properties of the AFXY on the UJ and the 
triangular lattice will be different, but to the best of our knowledge, there are no numerical researches investigated from the perspective of 
conformal field theory (CFT) on the AFXY on the UJ lattice.

Recently, tensor network (TN) algorithms have become a promising toolbox to investigate the criticality of frustrated systems 
\cite{PhysRevResearch.3.013041,PhysRevB.105.134516,PhysRevE.104.024118,PhysRevB.108.224404}. In this study, we extract the conformal data 
of the AFXY on the UJ lattice by applying recently proposed TN algorithms, the so-called tensor network renormalization (TNR) method 
\cite{PhysRevB.80.155131,PhysRevLett.118.110504,PhysRevLett.115.180405,PhysRevB.97.045111,homma2023nuclear}.  We believe that the main advantage of using TNR 
over the other existing methods is that one could improve the identification of the universality class, since this approach allows us to estimate not only the scaling dimensions of 
primary fields, but also a variety of conformal data, including central charges, higher levels of the scaling dimension spectrum, and operator product expansion (OPE) coefficients. 
In this study, we utilize one of the most sophisticated algorithms, so called nuclear norm regularized (NNR)-TNR algorithm, to elucidate the critical properties. 
Compared to other TNR algorithms, the NNR-TNR algorithm could potentially allow us to extract much more stable conformal data at larger system sizes with smaller 
bond dimensions. We expect this feature to be particularly effective in circumstances where multiple critical points are close together, 
as it allows us to suppress the notorious finite-size effect of critical phenomena. However, due to limited computational resources, 
it is always preferable to represent the TN states in a compact manner.

\begin{figure}
 \includegraphics[scale=0.42]{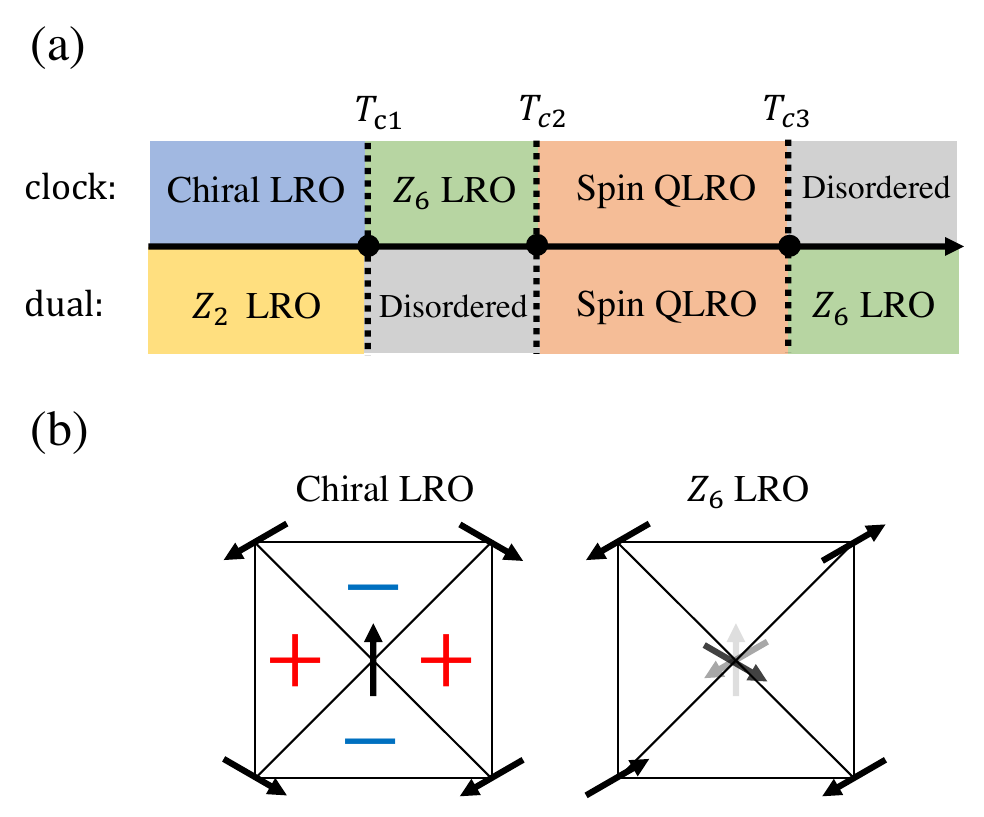}
 \centering
 \caption{\label{fig:spin_config}(a) (Top) Phase diagram of the AF 6-state clock model on the UJ lattice. With decreasing temperature the phase becomes disordered, 
 spin QLRO, $Z_{6}$ and chiral LRO, leading to the BKT, the $Z_{6}$ symmetry breaking and the Ising transition of the $Z_{2}$ 
 symmetry breaking of 6-fold degenerate states at $T_{\text{c3}}$, $T_{\text{c2}}$ and $T_{\text{c1}}$, respectively. 
 (Bottom) Phase diagram of the AF 6-state clock model in the dual representation. After the duality transformation, the $Z_{6}$ LRO phase is exchanged with the disordered phase, 
 while the chiral LRO phase is transformed into the $Z_{2}$ LRO phase. The corresponding universality classes at $T_{\text{c2}}$, $T_{\text{c3}}$ are also 
 swapped to the BKT and $Z_{6}$ symmetry breaking transition, respectively. 
 It exhibits the standard Ising transition at $T_{c1}$ via the $Z_{2}$ symmetry breaking. 
 (b) (Left) An example of spin configuration of the AF 6-state clock model at the ground state. The plus-minus signs indicate a chirality of the UJ plaquettes. 
 (Right) Its typical spin configuration of the $Z_{6}$ LRO phase \cite{PhysRevB.96.144432}. The fluctuating spin in the middle represents the disordered state.}
\end{figure}

In order to accurately determine the critical properties of the AFXY model on the UJ lattice, we first consider the discretized version of the continuous degrees of freedom, 
namely the AF 6-state clock model on the UJ lattice defined by the reduced Hamiltonian as,
\begin{eqnarray}
 \label{eq.1}
 H = J \sum_{\langle i,j\rangle}\cos(\theta_{i}-\theta_{j}),
\end{eqnarray}
where $\theta_{i} = 2\pi n_{i}/6$ and $n_{i}= 0,1\cdots 5$. The geometrical frustration is induced by the AF coupling on the UJ lattice and is expected to have $\pm 2\pi /3$ degree rotated 
spin configurations at the ground state with 12-fold degeneracy, as shown as “chiral LRO" in Fig.~\ref{fig:spin_config}(b). In this way, this model still remains as commensurate with the AFXY 
model on the UJ lattice, while suppressing the finite bond dimension effects of TN algorithms. For example, thanks to the discretization of the XY spins, the infinite ground state degeneracy of 
the AF 6-state clock model is alleviate to 12-fold degeneracy. In the case where the phase transition can be attributed to symmetry breaking, the finite number of ground state degeneracy enables us 
to track the critical properties at finite bond dimension via the TNR algorithm. 

In addition, since this discretization allows us to perform the Kramers-Wannier duality transformations \cite{PhysRev.60.252, Wu1976DualityTI,PhysRevB.16.1217}, 
we gain insight into the dual representation of the AF 6-state clock model.
Since the dual representation is nothing more than an alternative TN expression of the partition function defined on the dual lattice,
it should not affect the physical observables or the transition temperatures.
However, its phase behavior and the underlying scaling dimension may appear different from the original TN representation,
because its scaling field is also redefined in the dual representation. 
We observe that the dual TN representation leads not only to a further reduction of the required bond dimensions, 
but also to a simplification of the underlying conformal data.

We numerically confirm the phase diagram shown in Fig.~\ref{fig:spin_config}(a), as we will see in the rest of this article. The NNR-TNR calculation shows that the 
BKT, $Z_{6}$ symmetry breaking and Ising transition occur at different temperatures. For the AF 6-state clock model on the UJ lattice, as shown at the top of 
Fig.~\ref{fig:spin_config}(a), the lower and intermediate temperature phases correspond to the chiral and $Z_{6}$ LRO  (See also Fig.~\ref{fig:spin_config}(b)), while it exhibits the spin QLRO and 
disordered phase with increasing temperature.
The corresponding phase transitions are identified as the Ising, $Z_{6}$ symmetry breaking and BKT transition at $T_{\text{c1}}, T_{\text{c2}}$ and $ T_{\text{c3}}$, respectively. 
In the case of the dual representation, as shown at the bottom of Fig.~\ref{fig:spin_config}(a), the order of the $Z_{6}$ LRO phase and the disordered phase is swapped, but the chiral 
LRO phase is translated into the $Z_{2}$ ordered phase in the dual representation. Since the scaling field is rewritten in the dual way, the universality classes at $T_{\text{c1}}, T_{\text{c2}}$ 
and $ T_{\text{c3}}$ in the dual representation are converted to the Ising, BKT and $Z_{6}$ symmetry breaking transition, respectively. More precisely, the low-energy physics of both representations 
in the spin QLRO regime is consistent with the compactified boson CFT, while the universality classes at $T_{\text{c2}}$ and $T_{\text{c3}}$ are swapped after the duality transformation, 
which follows from the duality of the $Z_{q=6}$ deformed sine-Gordon theory. Furthermore, we confirm that the critical fixed point of the tensor at the $Z_{2}$ symmetry breaking point is in 
precise agreement with the Ising CFT, including central charge, scaling dimension spectrum and OPE coefficients.

This paper is organized as follows. In the Sect.~\ref{Tensor network}, we explain the recipes for the TN representations of the partition functions of the AF 6-state clock model 
and its dual representation, and the method to extract the physical quantities via the NNR-TNR. The Sect.~\ref{results} shows our main numerical results on the identification of 
their critical properties. Lastly, we give our conclusion and discussion of our work in Sect.~\ref{disc}.

\section{Tensor Network (TN)}
\label{Tensor network}
\subsection{TN construction of partition function of the AF 6-state clock model}
\label{sec2_a}
 The partition function of the classical Hamiltonian can be represented by trace of local tensors $\tau$ in the TN formalism as follows,
\begin{eqnarray}
 \label{eq.4}
 Z = \trace e^{\beta H } = \mathrm{tTr} \bigotimes \tau,
\end{eqnarray}
where $\beta=1/T$ is the inverse temperature and $ \mathrm{tTr}$ denotes trace over TNs.
As has been discussed in the previous literatures \cite{PhysRevB.105.134516}, the local tensor constructed under the standard TN formalism does not allow us to calculate the physical observable 
at thermodynamic limit. In fact, there are suitable choices of local tensors $\tau$ for frustrated spin systems. The gists of the proper choice can be rephrased in terms of encoding emergent degrees 
of freedom and constructing sets of local Hamiltonian tiles over whole lattice \cite{PhysRevResearch.3.013041,PhysRevB.108.224404}. We observed that this essence is passed on to the TNR algorithm, 
where the algorithm involves truncation of local tensors.
\begin{figure}
 \includegraphics[scale=0.42]{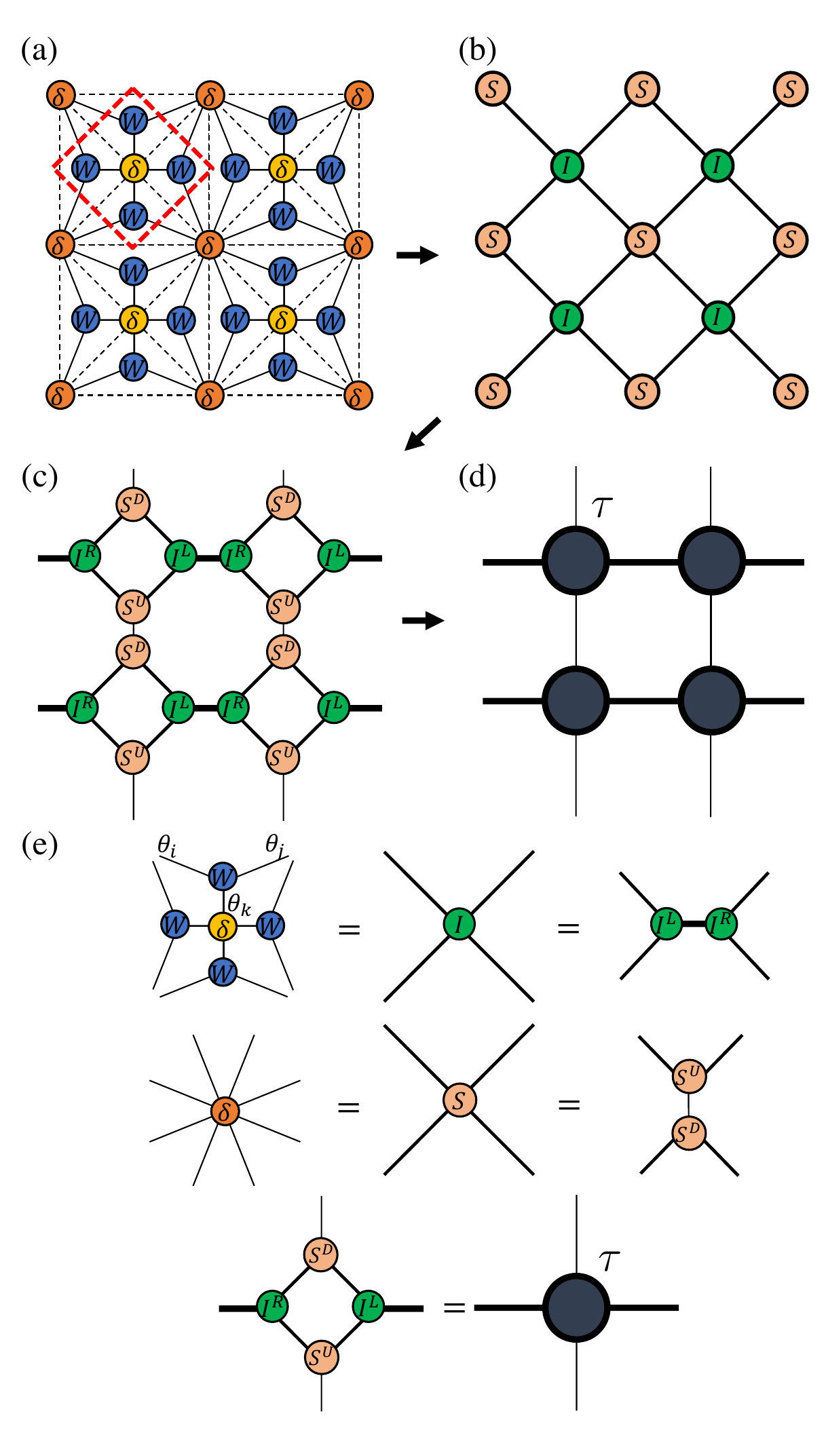}
 \centering
 \caption{\label{fig:original_lattice}(a) TN construction of partition function of the AF 6-state clock model on the UJ lattice. The dashed lines denote the original UJ lattice and the solid lines correspond to the connecting legs of TN. The $\delta$ tensor denotes the Dirac delta function and $W$ corresponds to the Boltzmann weight. (b) Mapping of the original UJ lattice to the TN on the square lattice by regrouping the indices of the tensor. (c) TN configurations after the decomposition of the $I$ and $S$ tensor. (d) Uniform TN representation of partition function on the square lattice. (e) The detailed procedures performed in the (a)-(d) for the construction of the local tensor.}
\end{figure}

The appropriate choice of the local tensors $\tau$ can be found via defining the Boltzmann weight of spin configurations on the dual site to include the emergent degrees of freedom. It is straightforward to define such a Boltzmann weight tensor (Fig.~\ref{fig:original_lattice}(a)) on the dual site as,
\begin{eqnarray}
 \label{eq.5}
 W(\theta_{i},\theta_{j},\theta_{k})= e^{\frac{J\beta}{2}[\cos(\theta_{i}-\theta_{j})+\cos(\theta_{j}-\theta_{k})+\cos(\theta_{k}-\theta_{i})]},
\end{eqnarray}
where the factor $1/2$ is introduced to avoid double counting.

The next step is to arrange the $\delta$ tensor so that the individual clusters are properly connected, i.e. the shared spins between different clusters should be consistent with each other. This can be achieved by placing the $\delta$ tensor on the original lattice vertices, but with its legs connected to the Boltzmann weight $W$. To illustrate this, we first split the UJ lattice into the two independent sublattices $A$ and $B$ and label them $A_{8}$ and $B_{4}$. Here their subscripts denote their coordination number, and each sublattice $A_{8}$ and $B_{4}$ is shown as yellow and orange circles in Fig.~\ref{fig:original_lattice}(a). The $\delta$ tensor on the $B_{4}$ sublattice can be defined as,
\begin{eqnarray}
 \label{eq.6}
 \delta_{\theta_{i},\theta_{j},\theta_{k},\theta_{l}} = 
 \left\{
 \begin{array}{ll}
 1, & \theta_{i} = \theta_{j} = \theta_{k} = \theta_{l}\\
 0, & \text{otherwise}.
 \end{array}
 \right.
 \end{eqnarray}
 The $\delta$ tensor on the $A_{8}$ sublattice can be constructed in a similar fashion as shown in Fig.~\ref{fig:original_lattice}(a).

 Let us first consider mapping of the tensor objects onto the square lattice shown in Fig.~\ref{fig:original_lattice}(b) for the convenience of the TNR algorithm. The Boltzmann weight $W$ and the $\delta$ tensor on the $B_{4}$ sublattice are regrouped as circled in the red dashed line of Fig.~\ref{fig:original_lattice}(a) to form a 4-leg tensor denoted by $I$.
 For the $\delta$ tensor on the $A_{8}$ sublattice, we simply regroup the two legs of the $\delta$ to form the 4-leg tensor denoted by $S$; see Fig.~\ref{fig:original_lattice}(e).
 
Lastly, in order to obtain the uniform TN representation of the local tensor $\tau$, we decompose the 4-leg tensors $I$ into the product of two 3-leg tensors horizontally in Fig.~\ref{fig:original_lattice}(c) via truncated singular value decomposition (SVD) as
 \begin{eqnarray}
 \label{eq.7}
 I&= U \Sigma V^{\dagger}
 &= I^{L}I^{R},
 \end{eqnarray}
 where $U$ and $V$ are 3-leg isometries and $\Sigma$ is a diagonal matrix. 
 Here we keep all singular values $\Sigma$ as long as they are greater than the machine epsilon and split it into two 3-legged tensors as $I^{L} = U\sqrt{\Sigma}$ and $I^{R} = \sqrt{\Sigma}V^{\dagger}$.
 The 4-leg tensor $S$ can be split vertically into a pair of 3-leg tensors $S^{U}$ and $S^{D}$ using the same procedures.
 Note that the thicker the legs of tensors, the larger bond dimensions required for the truncated SVD. Finally, contracting the 3-leg tensors in Fig.~\ref{fig:original_lattice}(e) gives the uniform TN representation of the local tensor $\tau$ in Fig.~\ref{fig:original_lattice}(d).
 
\subsection{TN construction of partition function in the dual representation}
\label{sec2_b}
In this section, we consider mapping the original model onto the dual lattice via the Kramers-Wannier duality transformation. 
This approach has several advantages; (1) the connectivity of TN can be reduced, while the degrees of freedom for each index are rather compact. 
This allows to mitigate the effects of finite bond dimensions. (2) The ground state degeneracies of the given model could be reduced in the dual representation.
\begin{figure}
 \includegraphics[scale=0.42]{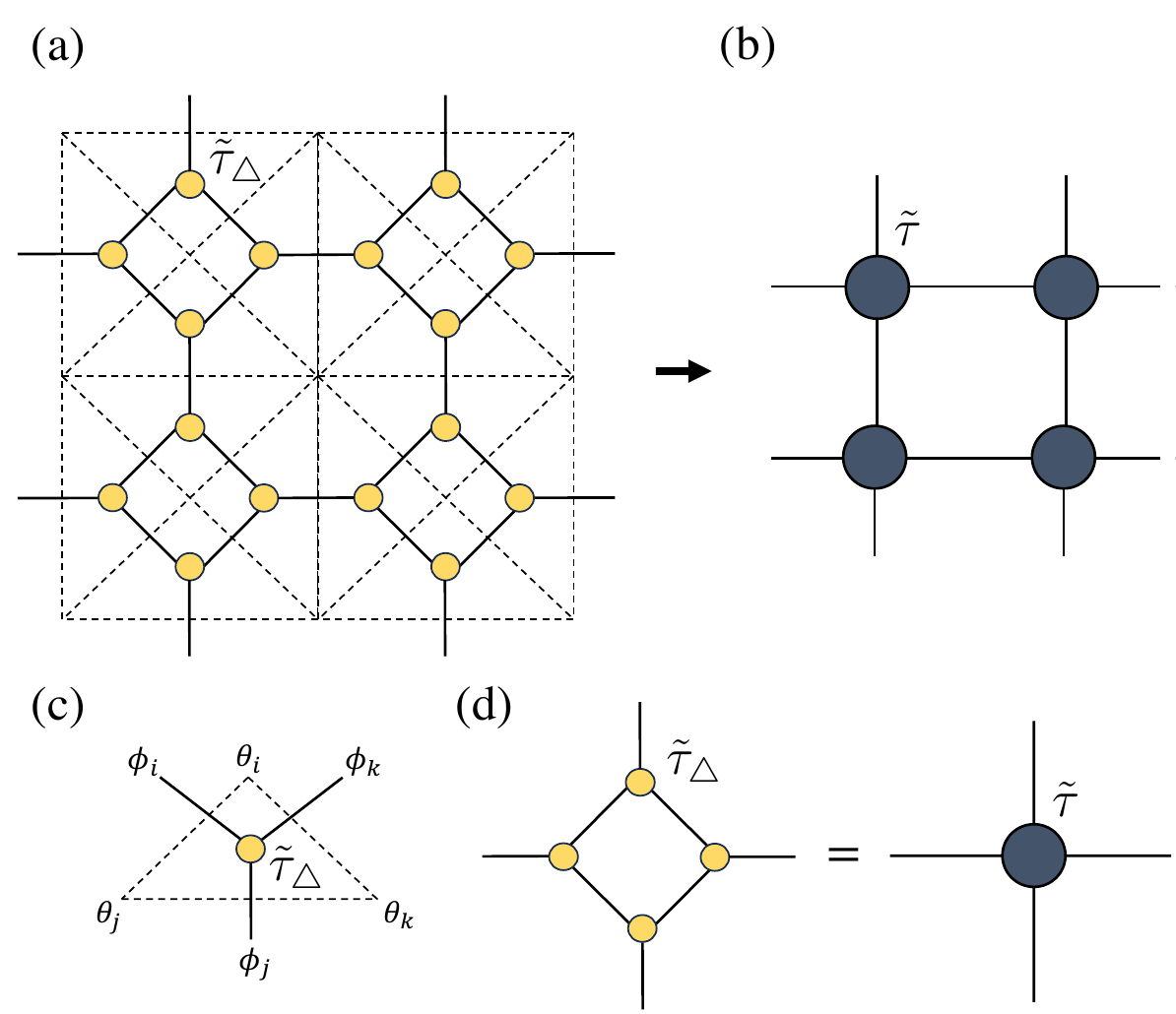}
 \centering
 \caption{\label{fig:dual_lattice}(a) TN construction of partition function in the dual representation. The 3-leg tensor $\tilde{\tau}_{\bigtriangleup}$ is defined on the dual lattice. Again, the black dashed line corresponds to the original UJ lattice. (b) One can map the dual TN representation onto the square lattice by grouping the 3-leg tensors $\tilde{\tau}^{\bigtriangleup}$. (c) The original and the corresponding dual variables defined on the triangular plaquette. (d) Summation over the square gives the local tensor $\tilde{\tau}$.}
 \end{figure}
To illustrate this, let us define new variables in the dual lattice \cite{PhysRevResearch.4.023159,PhysRevE.101.060105,Chen_2017}, as shown in Fig.~\ref{fig:dual_lattice}(c),
\begin{eqnarray}
 \label{eq.8}
 \phi_{i} = \theta_{i}-\theta_{j},\nonumber\\
 \phi_{j} = \theta_{j}-\theta_{k},\nonumber\\
 \phi_{k}= \theta_{k}-\theta_{i}.\nonumber
\end{eqnarray}
These three dual variables form the dual lattice in Fig.~\ref{fig:dual_lattice}(a). In this dual representation, it is reduced to a loop model without the closed loops conditions, 
i.e. the dual variable $\phi$ may form closed loops or fluctuating strings with two point charges.
The dual TN representation of a small patch of the local tensor $\tilde{\tau}_{\bigtriangleup}$ is given by

\begin{eqnarray}
 \label{eq.9}
 \tilde{\tau}_{\bigtriangleup} = \frac{1}{\sqrt{6}}e^{-\frac{J\beta}{2}[\cos(\phi_{i})+\cos(\phi_{j})+\cos(\phi_{k})]}\delta^{(6)}_{i+j+k},
\end{eqnarray}
where $\delta^{(6)}_{i+j+k}$ denotes $\delta_{[\text{mod}(\phi_i + \phi_j + \phi_k,6) ,0]}$. 
Again, one can obtain the uniform TN representation of the local tensor $\tilde{\tau}$ by mapping the TN states arranged on the dual lattice 
onto the square lattice (Fig.~\ref{fig:dual_lattice}(b)) by contracting the 3-leg tensor $\tilde{\tau}_{\bigtriangleup}$ as shown in Fig.~\ref{fig:dual_lattice}(d).
The general construction of the dual TN representations is discussed in Appendix~\ref{dualTN}.

It should be noted that the dual TN representation of the local tensor $\tilde{\tau}$ constructed above requires much 
fewer bond dimensions than the original local tensor $\tau$. This is due to the fact that its dual lattice has less connectivity compared 
to the original UJ lattice. In addition, the duality transformation serves as reduction of ground state degeneracies. 
To see this, we remind the reader that the physical pictures of the dual representation are represented by spin difference $\theta_{i}-\theta_{j}$ 
of the original model. For example, the chiral LRO has 12-fold degeneracy with $\pm2\pi /3$ degree spin differences. 
In the dual representation, the dual variable behaves like Ising spins as $\cos(\phi)= \pm 1/2$ and leads to a reduction of 12-fold ground state degeneracy to twofold, 
namely the $Z_{2}$ LRO phase. Similarly, the $Z_{6}$ LRO phase is converted to the trivial state as disordered state, since the dual variable always gives $\cos(\phi)= -1$. 
The concrete benefit of the reduction of ground state degeneracy for the TN method will be discussed in the Sect.~\ref{results}.
\subsection{Calculation of physical quantities}
To calculate the expectation value of the physical quantities $O$, such as internal energy and chirality, we perform the impurity method with loop optimization (see Appendix ~\ref{imp} for details).
In this particular study, we split the raw interactions on the UJ lattice into the two independent sublattices $A_{8}$ and $B_{4}$. Since this definition cannot be realized on the dual lattice, the impurity tensors after the duality transformation do not appear in this work. 

The impurity tensor corresponding to internal energy $E$ of the AF 6-state clock model on the UJ lattice can be constructed by replacing the Boltzmann weight $W$ with the following,
\begin{equation}
  \begin{aligned}[b]
  \label{eq.102}
  W^{E}= [\cos(\theta_{i}-\theta_{j})+\cos(\theta_{j}-\theta_{k})+\cos(\theta_{k}-\theta_{i})]\times W.
 \end{aligned}
 \end{equation}
We measure the internal energy $E$ on each sublattice $A_{8}$ and $B_{4}$ separately and estimate the specific heat $C_{A_{8},B_{4}}$ from the numerical derivatives of $dE/dT$.

To discuss the $Z_{2}$ symmetry of the AF 6-state clock model, it is common to introduce an order parameter called chirality \cite{PhysRevE.66.026111,Tasrief_Surungan_2004}, which is defined by
\begin{equation}
 \label{eq.2}
 \begin{aligned}[b]
 \kappa = \frac{2}{3\sqrt{3}N_{\bigtriangleup}} \sum_{\bigtriangleup} [\sin(\theta_{i}-\theta_{j}) +\sin(\theta_{j}-\theta_{k}) + \sin(\theta_{k}-\theta_{i})],
 \end{aligned}
\end{equation}
where the indices $(i,j,k)$ run over the $N_{\bigtriangleup}$ triangular subplaquettes and the order of indices is chosen to be alternately clockwise and counterclockwise on the triangular subplaquettes.
The corresponding impurity tensor for chirality $\kappa$ can be defined as,
\begin{equation}
 \label{eq.10}
 \begin{aligned}[b]
 W^{\kappa}= [\sin(\theta_{i}-\theta_{j})+\sin(\theta_{j}-\theta_{k})+\sin(\theta_{k}-\theta_{i})]\times W.
 \end{aligned}
 \end{equation}

 For two-dimensional critical systems, the CFT serves as a powerful tool for understanding universality classes \cite{CFTtextbook}. The corresponding conformal data, including central charge, scaling dimension spectrum, and OPE coefficients, can be efficiently computed using NNR-TNR. For example, the Gu and Wen method \cite{PhysRevB.80.155131} allows us to estimate central charge and scaling dimension spectrum via diagonalizing the transfer matrix $M_{l_{x}}$ defined as below,
\begin{equation}
 \label{eq.11}
 \begin{matrix}
 \vspace*{0.6cm}
 \includegraphics[scale=0.28]{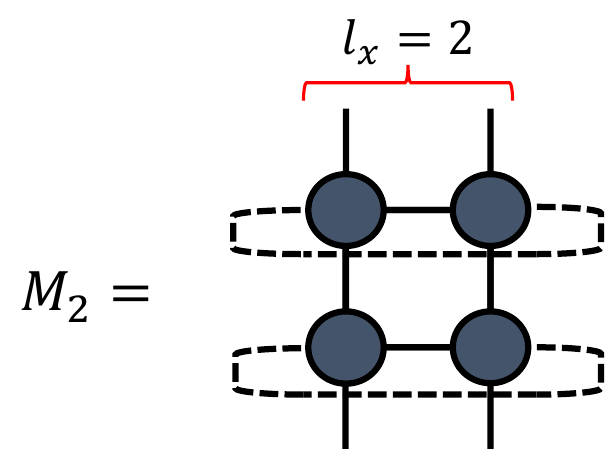} 
 \end{matrix}
 \ \rm{and} \ 
 \begin{matrix}
 \vspace*{0.6cm}
 \includegraphics[scale=0.28]{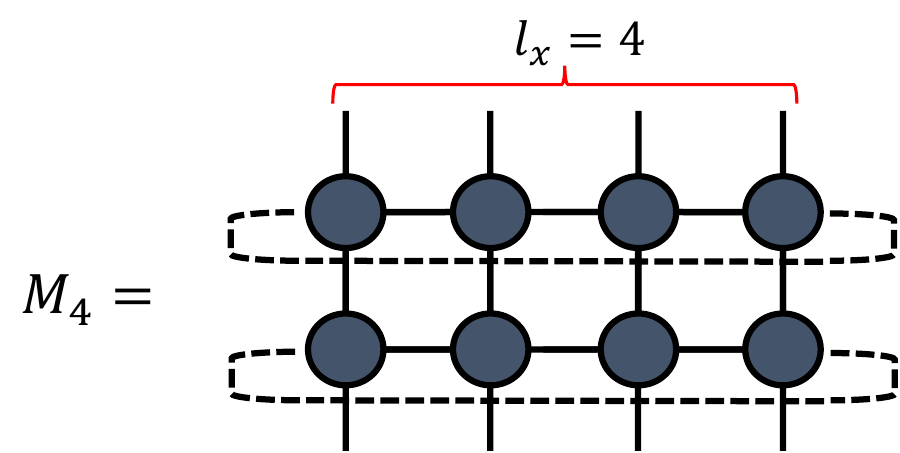}
 \end{matrix},
\end{equation}
Assuming that the lowest scaling dimension is $\Delta_{0}=0$, one can compute central charge and scaling dimension spectrum as 
\begin{eqnarray}
 \label{eq.144}
 c =\frac{6l_{x}}{\pi}(\ln \lambda_{1}+l_{x}N_{f}), \\
 \Delta_{\alpha}= -\frac{l_{x}}{2\pi}(\ln \lambda_{\alpha}-\ln \lambda_{1}).
\end{eqnarray}
Here, $l_{x}$ denotes the column length of the linear transfer matrix, $\lambda_{\alpha}$ correspond to the eigenvalues of the linear transfer matrix, and $N_{f}$ is a normalization factor.
In general, the larger column length $l_{x}$ leads to the reduction of finite-size effects, thereby enhancing accuracies. However, this improvement comes at the expense of increasing computational complexity. To mitigate this problem, we approximate the transfer matrix $M_{l_{x}}$ and diagonalize it via the locally optimal block conjugate gradient (LOBCG) \cite{doi:10.1137/S1064827500366124} method (see Appendix~\ref{lobcg} for details).

Furthermore, one can extract the OPE coefficients $\mathcal{C}_{\alpha \beta \gamma} $ from the overlaps of the eigenvectors of the transfer matrices \cite{PhysRevB.108.024413,PhysRevB.105.125125,PhysRevB.107.155124} as,
\begin{equation}
 \label{eq.145}
 \mathcal{C}_{\alpha \beta \gamma} = \frac{ \bra{\psi_{\alpha}(M_{4})} \ket{\psi_{\beta}(M_{2}) \psi_{\gamma}(M_{2})} }{\bra{\psi_{1}(M_{4})} \ket{\psi_{1}(M_{2})\psi_{1}(M_{2})}} 2^{-\Delta_{\alpha}+2\Delta_{\beta}+2\Delta_{\gamma}},
\end{equation}
where $\ket{\psi_{\alpha}(M_{l_{x}})}$ is the eigenvector of the transfer matrix $M_{l_{x}}$ corresponding to its eigenvalue $\lambda_{\alpha}$. For example, the Ising CFT is constructed from three primary operators; identity operator $\mathbf{1}$, spin operator $\sigma$, and thermal operator $\epsilon$. It is known that the OPE coefficients for the Ising CFT are
\begin{eqnarray}
 \label{eq.146}
 \mathcal{C}_{\sigma \sigma \mathbf{1} } = \mathcal{C}_{\epsilon \epsilon \mathbf{1}} = 1, \\
 \mathcal{C}_{\epsilon \sigma \sigma} = 1/2.
\end{eqnarray}
Note that although this approach is remarkably simple and convenient to extract the OPE coefficients, we find it very difficult to estimate the stable OPE coefficients on the critical region with $c=1$, which we believe is due to the truncation error of the NNR-TNR algorithm. In this study, we limit ourselves to apply this method only to the Ising CFT.

Finally, in order to detect the phase transition, the following gauge invariant quantity $X$ is convenient to locate the off-critical phases \cite{PhysRevB.80.155131}. In this study, we define the quantity $X$ as follows,
\begin{eqnarray}
 \label{eq.12}
 X = \frac{(\trace M_{2})^{2}}{\trace M_{4}}.
\end{eqnarray}
It is known that this quantity $X$ corresponds to the degeneracies of characteristic states.

\section{Numerical results}
\label{results}
\begin{figure}
 \includegraphics[scale=0.42]{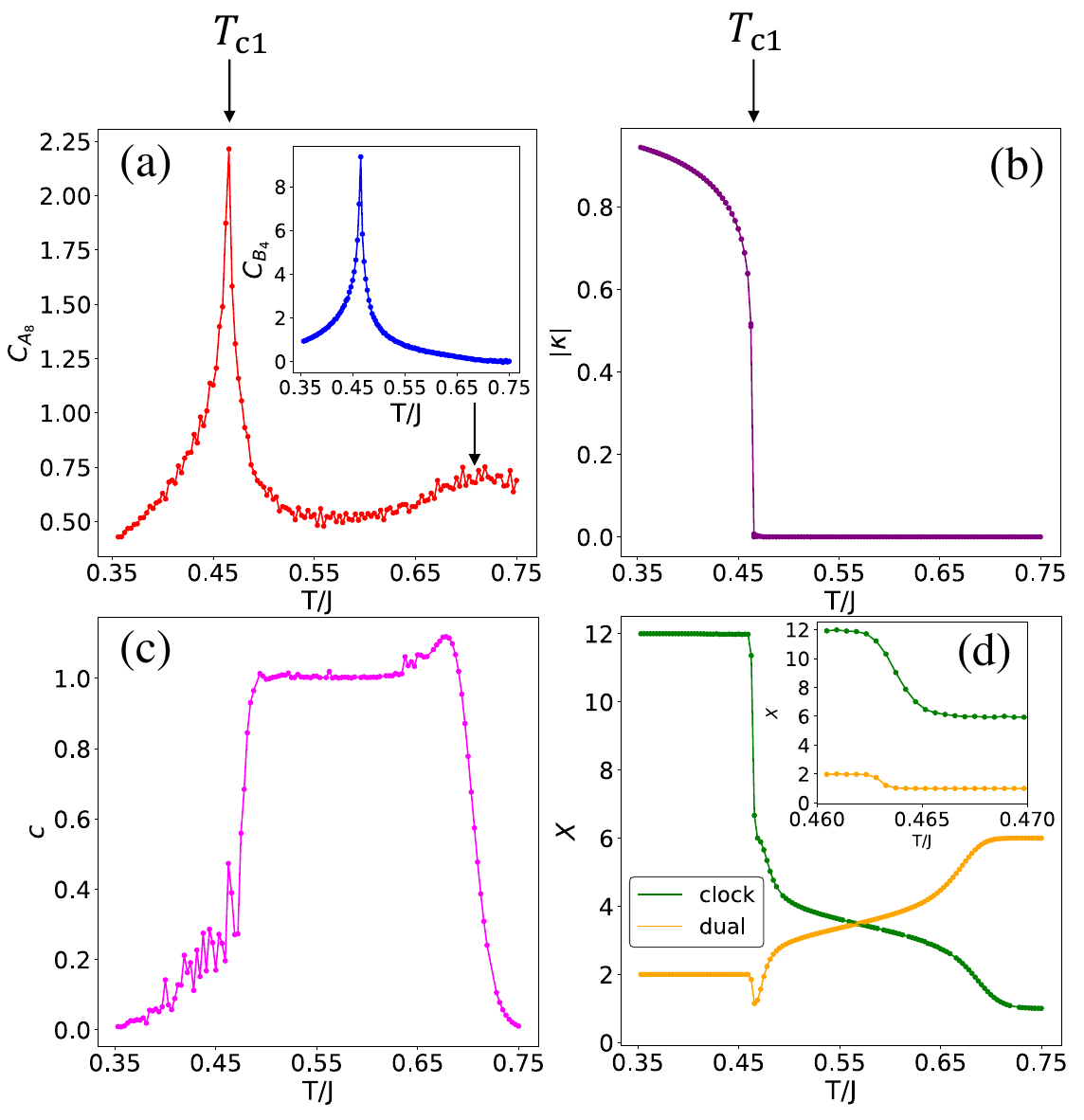}
 \centering
 \caption{\label{fig:physical_quantity_all} Physical quantities of the AF 6-state clock model as a function of the temperature: (a) the specific heat $C$ on the sublattice $A_{8}$. Its inset shows the specific heat on the sublattice $B_{4}$. (b) the absolute value of spin chirality $|\kappa|$. (c) central charge $c$. (d) Temperature dependence of gauge invariant quantity $X$ for the AF 6-state clock model and its dual representation. The inset shows the enlarged plot of the gauge invariant quantity $X$ around the Ising transition. These results are obtained by NNR-TNR with bond dimension $\chi = 36$ at the 50th RG step for the (a)-(b), and at the 12th RG step for the (c)-(d).}
\end{figure}
In this section we present some numerical results on the physical quantities and the conformal data of the AF 6-state clock model and its dual representation using the 
NNR-TNR algorithm\footnote{The hyperparameters of NNR-TNR defined in Ref.\cite{homma2023nuclear} have been chosen as $\xi = 2\cdot 10^{-6}$ and $\rho =0.9$ throughout the main context.}. 
Note that while the physical quantities such as specific heat $C$, chirality $\kappa$, central charge $c$ of the AF 6-state clock model and its dual representation should be equivalent at 
the thermodynamic limit, its gauge invariant quantities $X$ and the scaling dimension spectrum may mark different values. For those invariant quantities under duality transformation, 
we show only the results of the AF 6-state clock model, while both plots on its gauge invariant quantities and the scaling dimension spectrum
are included in the following.

To capture the essence of the phase transitions, we first plot the specific heat on two sublattices $C_{A_{8}}$ and $C_{B_{4}}$ of the AF 6-state clock model in 
Fig.~\ref{fig:physical_quantity_all}(a). For the specific heat on the sublattice ${A_{8}}$, there are a divergent peak at $T_{\rm{c1}}$ and a small bump at higher temperature, 
indicating the second order phase transition and other phase transitions such as the BKT transition, respectively. On the other hand, the specific heat on the sublattice $C_{B_{4}}$ 
shown in the inset of Fig.~\ref{fig:physical_quantity_all}(a) has only a single divergent peak. This is consistent with the previous study of the AFXY on the UJ lattice \cite{PhysRevB.87.024108}.
As shown in Fig.~\ref{fig:physical_quantity_all}(b), the local chirality $\kappa$ of the AF 6-state clock model vanishes around the chiral transition temperature $T_{\rm{c1}}$, 
indicating the formation of a chiral LRO phase. This again suggests a conventional second order phase transition with an order parameter like the Ising transition. 
From the singular point of the local chirality we estimate the chiral transition temperature to be $T_{\rm{c1}} = 0.462J$.

To identify the nature of the universality class, we plot central charge as a function of temperature in Fig.~\ref{fig:physical_quantity_all}(c). It appears to have $c \approx 1$ for the middle 
region and $c = 0.497$ at the lower temperature, consistent with the compactified boson and Ising CFT, respectively.

For the gauge invariant quantity $X$ of the AF 6-state clock model in Fig.~\ref{fig:physical_quantity_all}(d), the fixed-point tensor flows into the tensor with $X=1$ in the high temperature 
(disordered) phase, while it takes $X = 12$ in the low temperature (chiral LRO) phase, which corresponds to the ground state degeneracies of this model. Note that there is a sudden change in $X$ between 
the chiral LRO phase and the intermediate phase ($X=6$) around the Ising transition. This implies not only that the intermediate phase corresponds to the $Z_{6}$ LRO phase, but also that this Ising 
transition can be attributed to the $Z_{2}$ symmetry breaking of 6-fold degenerate states, yielding a total of 12-fold ground states. On the other hand, the gauge invariant quantity $X$ from the dual 
representation in Fig.~\ref{fig:physical_quantity_all}(d) suggests that the chiral LRO, $Z_{6}$ LRO and disordered phases are now translated into $Z_{2}$ LRO, disordered and $Z_{6}$ LRO phases, 
respectively. We observe that the sudden jump around the Ising transition is reduced from $X=6 \leftrightarrow 12$ to $X=1 \leftrightarrow 2$ in the inset of Fig.~\ref{fig:physical_quantity_all}(d) 
and is consistent with the property of the standard Ising universality class with $Z_{2}$ symmetry breaking.

\begin{figure}
 \includegraphics[scale=0.42]{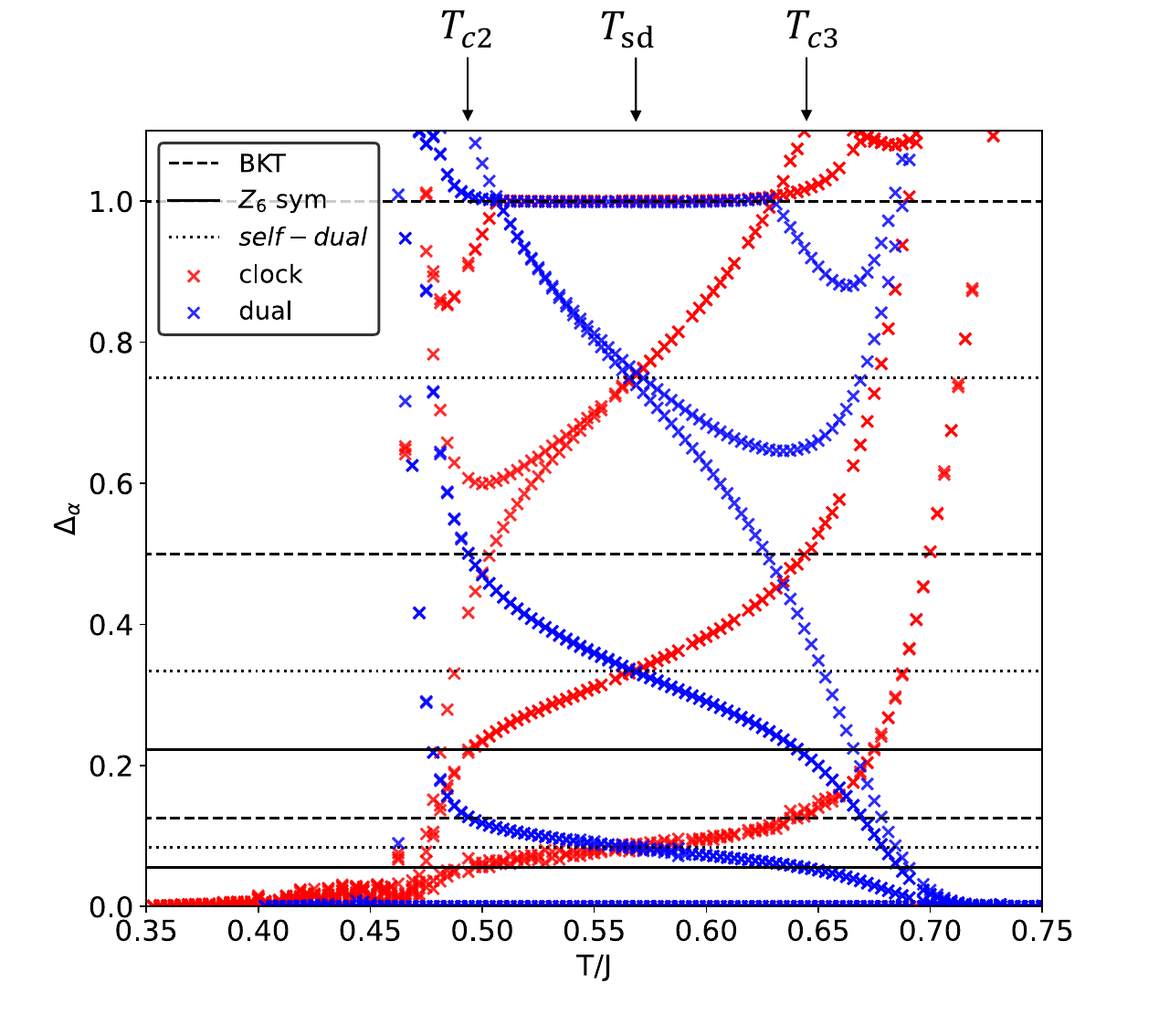}
 \centering
 \caption{\label{fig:dual_scaling_dims_KT} First 8 scaling dimensions as a function of temperature. These scaling dimensions are estimated at the 15th RG step with bond dimension $\chi =36$ using 
 the $l_{x}=2$ transfer matrix method. The black dotted, solid, and dashed lines correspond to the exact scaling dimensions of the BKT, $Z_{6}$ symmetry breaking, and self-dual point, respectively.} 
\end{figure}

In order to get an insight into the $c = 1$ region, 
we plot the first 8 scaling dimensions of the AF 6-state clock model and its dual representation as a function of temperature in Fig.~\ref{fig:dual_scaling_dims_KT}. 
We conjecture that the temperature dependence of scaling dimension spectra can be described by the $Z_{q=6}$ deformed sine-Gordon theory 
\cite{PBWiegmann_1978,PhysRevE.101.060105}
\begin{eqnarray}
\label{eq.131}
S &=& \frac{1}{2\pi K}\int d^{2}\bm{r}\nabla (\Phi)^{2} + \frac{g_{1}}{2\pi \alpha^{2}}\int d^{2}\bm{r} \cos(\sqrt{2}\Phi) \nonumber\\
&+& \frac{g_{2}}{2\pi \alpha^{2}}\int d^{2}\bm{r} \cos(q\sqrt{2}\Theta),
\end{eqnarray}
where $\Phi$ and $\Theta$ are real scalars and compactified as 
$\Phi \equiv \Phi+ \sqrt{2}\pi$, $\Theta \equiv \Theta+ \sqrt{2}\pi$. The coupling constants $K, g_{1}, g_{2}$ are a function of temperature. $\alpha$ is an ultraviolet cutoff.
The fields $\Phi$ and $\Theta$ are dual to each other by,
\begin{eqnarray}\label{eq.110}
\partial_{x}\Phi = -K\partial_{y}\Theta, \ \ \ 
\partial_{y}\Phi = K\partial_{x}\Theta. 
\end{eqnarray}

Within the critical region $T_{c2}< T <T_{c3}$, the first term in Eq.~\eqref{eq.131} is relevant and corresponds to the $c=1$ compactified boson CFT. 
It is well known that the scaling dimensions of the compactified boson can be expressed as
\begin{eqnarray}\label{eq.13}
 \Delta_{m,n}(K) = \frac{1}{2}\left(m^{2}K+\frac{n^{2}}{K}\right),
\end{eqnarray}
where $m,n$ are some integers. For its BKT transition, and $Z_{q}$ symmetry breaking transition, 
the corresponding coupling constants are $K_{\text{BKT}} = 4$ and $K_{Z_{q}} = q^{2}/4$, respectively.

In order to discuss the duality in the sine-Gordon theory, we consider the following dual transformation \cite{Matsuo_2006}
\begin{eqnarray}\label{eq.111}
\Phi = q\Theta, \ \ q\Theta= \Phi.
\end{eqnarray}
Replacing Eq.~\eqref{eq.131} by Eq.~\eqref{eq.111} results in
\begin{eqnarray}
  \label{eq.132}
  S_{\text{dual}} &=& \frac{q^{2}}{2\pi K}\int d^{2}\bm{r}\nabla (\Theta)^{2} + \frac{g_{1}}{2\pi \alpha^{2}}\int d^{2}\bm{r} \cos(q\sqrt{2}\Theta) \nonumber\\
  &+& \frac{g_{2}}{2\pi \alpha^{2}}\int d^{2}\bm{r} \cos(\sqrt{2}\Phi).
\end{eqnarray}
Substituting Eq.~\eqref{eq.110} into Eq.~\eqref{eq.132} gives the same form as Eq.~\eqref{eq.131} except for the coefficients. 
To achieve their correspondence, one must exchange the coupling constants as $K \leftrightarrow q^{2}/K$ and $g_{1} \leftrightarrow g_{2}$.
If $g_{1}=g_{2}$, it corresponds to the self-dual point of the effective field theory and should have the same scaling dimensions with $K_{\text{sd}}=q$.
Note that this self-dual point is not obvious from the lattice model or the raw Boltzmann weight $W$, 
but rather within the framework of effective field theory, i.e., the fixed-point tensor under the RG transformation. 
It should also be noted that in the $c=1$ region, the scaling dimension in Eq.~\eqref{eq.131} as a function of $K$ 
should be equivalent to the the scaling dimension in Eq.~\eqref{eq.132} as a rescaled function of $K^{2}_{\text{sd}}/K$.


The scaling dimension spectra of the AF 6-state model and its dual representation shown in Fig.~\ref{fig:dual_scaling_dims_KT} seem to intersect roughly at a single point 
$T_{\text{sd}}=0.56J$ with the predicted scaling dimension. This suggests the possibility of the self-dual point in the AF 6-state clock model.
We also find that the scaling dimension spectrum of the AF 6-state clock model at $T$ is in good agreement 
with that of the dual representation at the rescaled temperature $T^{2}_{\text{sd}}/T$, showing a fair correspondence in line with our conjecture.
From the points where the two lowest levels of scaling dimensions intersect the exact one with $K_{Z_{6}}$ and $K_{\text{BKT}}$,
we identify the universality classes at $T_{\text{c2}}$ and $T_{\text{c3}}$ of the AF 6-state model as the $Z_{6}$ symmetry breaking and the BKT transition, 
and estimate the corresponding transition temperatures to be $T_{\text{c2}}=0.49J$ and $T_{\text{c3}} =0.64J$, respectively.
Since the dual representation rewrites the field as $K \leftrightarrow q^{2}/K$ and $g_{1} \leftrightarrow g_{2}$,
the universality classes at $T_{\text{c2}}$ and $T_{\text{c3}}$ in the dual representation are reversed as the BKT and the $Z_{6}$ symmetry breaking transition, respectively. 

\begin{figure}
 \includegraphics[scale=0.41]{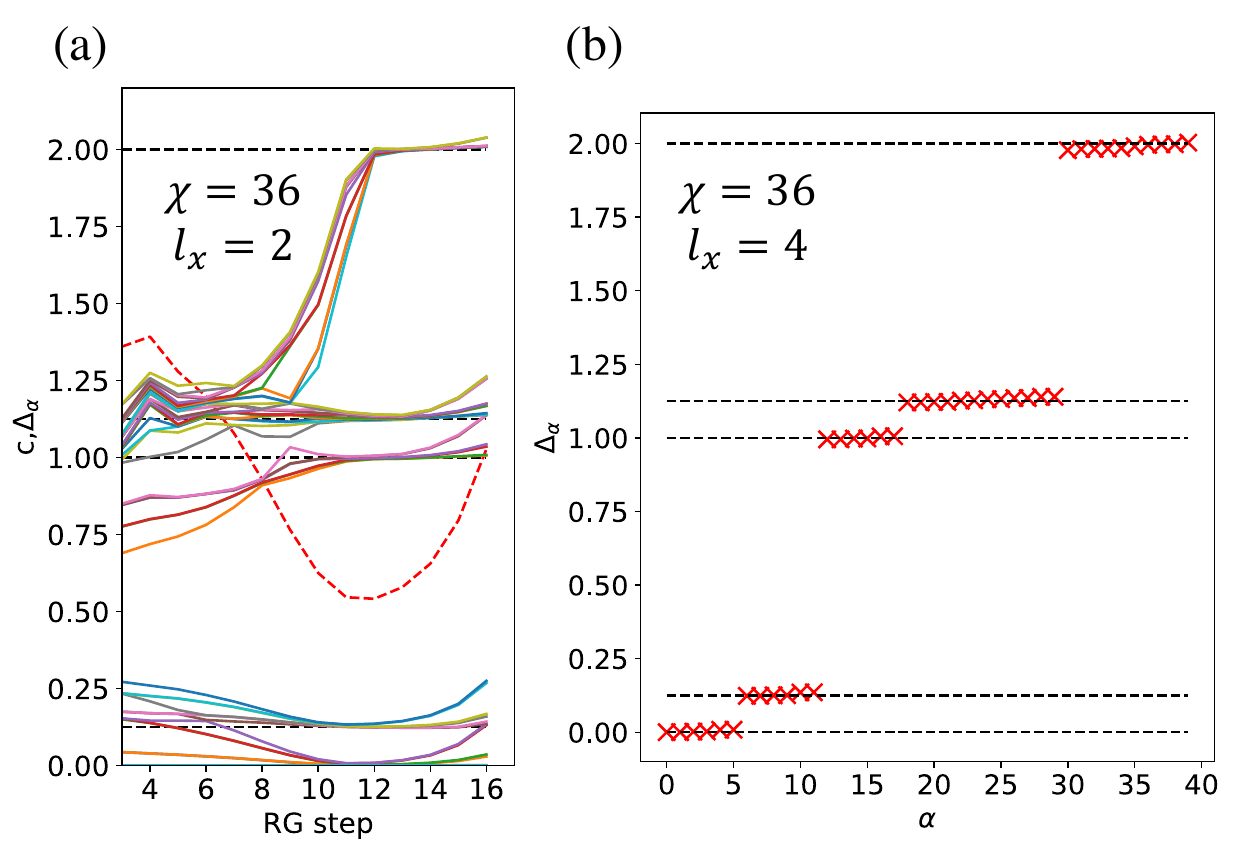}
 \centering
 \caption{\label{fig:UJ_Ising_spectrum} (a) Conformal data of the AF 6-state clock model as a function of different RG steps at the Ising transition temperature $T_{\text{c1}}$. The red dashed line 
 represents central charge and the other solid lines denote scaling dimension spectrum. (b) The scaling dimension spectrum as a function of index at the 12th RG step. The the black dashed lines 
 correspond to the exact scaling dimensions of the Ising CFT, while the red crossed points are estimated from the NNR-TNR algorithms. We extract these conformal data using the transfer matrix methods 
 with $l_{x}=2,4$ for the (a) and (b), respectively.
 }
\end{figure}
Fig.~\ref{fig:UJ_Ising_spectrum}(a) shows the conformal data of the AF 6-state clock model at the obtained chiral transition $T_{\rm{c1}}= 0.462J$ using the the $l_{x}=2$ transfer matrix method. For the early RG steps before the 12th RG step, we observed that the central charge along the RG steps deviates from $c=1/2$. This behavior is due to the fact that two transition temperatures are so close that the finite size effect induced by the nearby $Z_{6}$ symmetry breaking transition makes it difficult to obtain the true critical behavior from small systems. Around the 12th RG step, the extracted conformal data starts to collapse onto the flat and degenerate spectrum, indicating the Ising universality class without a prior knowledge of the conformal data. At the 12th RG step it roughly agrees with the Ising CFT, including central charges and the scaling dimensions $\Delta_{\alpha} <3$ in Fig.~\ref{fig:UJ_Ising_spectrum}(a), although the stability of the conformal data deviates from the expected value afterwards. In Fig.~\ref{fig:UJ_Ising_spectrum}(b), using the longer columns of the transfer matrix with $l_{x}=4$, we find that its scaling dimension spectrum is consistent with 6-fold degenerate Ising CFT, namely there are 6 copies of identity, spin and thermal operators due to the $Z_{2}$ symmetry breaking of 6-fold degenerate states.

\begin{figure}
 \includegraphics[scale=0.42]{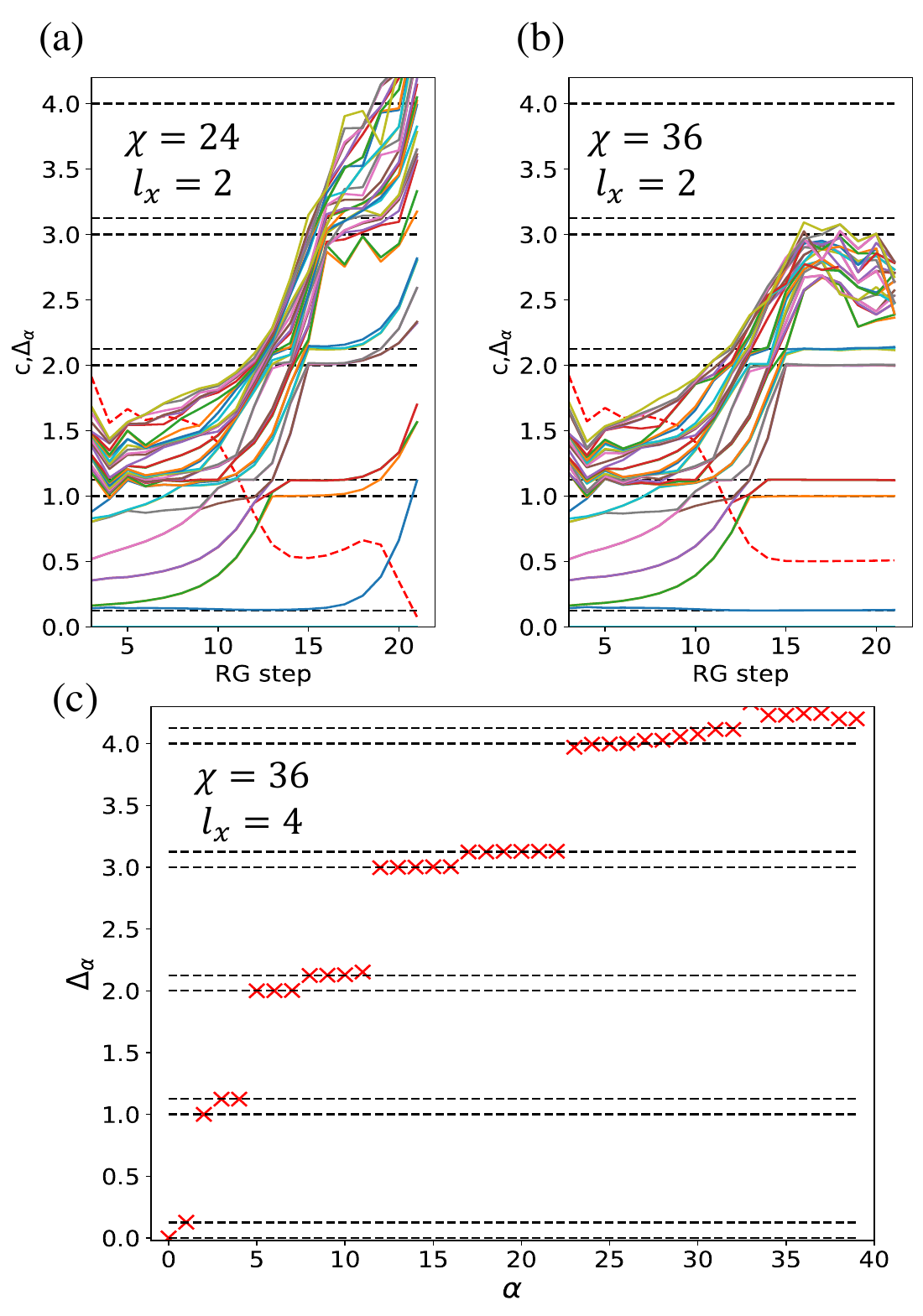}
 \centering
 \caption{\label{fig:CFT_TAFclock} (a) and (b): Conformal data from the dual representation as a function of different RG steps at the same Ising transition temperature $T_{\text{c1}}$ with different 
 bond dimensions $\chi =24,36$. (c) The scaling dimension spectrum as a function of index at the 20th RG step. We extract these conformal data using the transfer matrix methods with $l_{x}=2$ for 
 the (a)-(b), and $l_{x}=4$ for the (c).
 }
\end{figure}

Fig.~\ref{fig:CFT_TAFclock}(a) and (b) show the conformal data from the dual representation at the same temperature $T_{\rm{c1}}$ with bond dimension $\chi=24,36$, respectively. Our result is again deteriorated by the finite size effects of the nearby the BKT transition point for earlier RG steps, but eventually flows into the correct fixed-point tensor consistent with the Ising CFT for $\Delta_{\alpha}< 3$. We note that the stability of the conformal data along the RG steps in Fig.~\ref{fig:CFT_TAFclock}(b) is drastically improved from Fig.~\ref{fig:CFT_TAFclock}(a) with the larger bond dimensions. With the longer column length of transfer matrix with $l_{x}=4$ and the LOBCG method, we obtain that the higher levels of scaling dimension spectrum in Fig.~\ref{fig:CFT_TAFclock}(c) seem to be further improved with the longer column length of transfer matrix, i.e. the scaling dimensions $\Delta_{\alpha}<4$.
In addition, we estimate its OPE coefficients from the overlap of the eigenvectors of the transfer matrix $M_{l_{x}=2,4}$ using the dual representation at the 20th RG step as $\mathcal{C}_{\sigma \sigma \mathbf{1}}=1.046$, $\mathcal{C}_{\epsilon \epsilon \mathbf{1} }=1.003$ and $\mathcal{C}_{\epsilon \sigma \sigma}=0.529$, which are evidently close to the values expected for the Ising CFT in spite of the interference from the nearby BKT transition. Based on these conformal data computed above, it seems reasonable to argue that the chiral transition of the AF 6-state clock model on the UJ lattice belongs to the Ising CFT. The comparison between Fig.~\ref{fig:UJ_Ising_spectrum} and Fig.~\ref{fig:CFT_TAFclock} clearly shows the advantage of the dual representation, which reduces the bond dimension requirement by removing the 6-fold degeneracy in the original clock model.

\section{Conclusions and discussions}
\label{disc}
In summary, we have studied the phase diagrams and their critical properties of the AF 6-state clock model on the UJ lattice and its dual representation. Based on the NNR-TNR algorithm, 
we find that the AF 6-state clock model has multiple phase transitions, the BKT, $Z_{6}$ and Ising transition with decreasing temperature. Compared to the previous study on the the AFXY model 
on the UJ lattice, we have provided the more reliable classification of its critical properties based on the conformal data; the low-energy physics of the spin QLRO phase is well explained by 
the $c=1$ compactified boson theory and its relation to the dual representation is described by the duality of the $Z_{q=6}$ deformed sine-Gordon theory. In addition, we find the chiral 
transition is in good agreement with the Ising CFT. Interestingly, we observe that the Ising transition of the AF 6-state clock model consists of 6-fold degenerate scaling dimensions of the primary operators due to the $Z_{2}$ symmetry breaking of 
the 6-fold state degeneracy, while that of the dual representation is in perfect agreement with the standard Ising CFT, including the central charge, the scaling dimension spectrum and the OPE coefficients. 

Our result suggests that the TN method and the dual representation could serve as a powerful tool to obtain accurate conformal data in the frustrated system. In the present case, the duality 
transformation not only acts as a reduction of the necessary bond dimensions required to represent the local tensor, but also allows us to rewrite redundant copies of primary operators of 
the underlying CFT into a minimal representation. In particular, in scenarios where one seeks to estimate conformal data of continuous spin frustrated systems on the original lattice model, 
one can infer that due to the infinite ground state degeneracy, obtaining accurate conformal data within the TN method may be considered impossible. The combination of appropriate discretization 
and its dual representation with TN methods may open new avenues for understanding critical phenomena that have been difficult to accurately confirm in previous studies.
\begin{acknowledgments}
 We are grateful to Xinliang Lyu, Katsuya Akamatsu, Atsushi Ueda and Kenji Harada for fruitful discussion. S.M. is supported by the Center of Innovations for Sustainable Quantum AI (JST Grant Number JPMJPF2221) and by JSPS KAKENHI Grant Numbers JP20K03780 and JP23H03818. 
 N.K. is supported by JSPS KAKENHI Grants Numbers JP19H01809 and JP23H01092. The computation in this work has been done using the facilities of the Supercomputer Center, the Institute for Solid State Physics, the University of Tokyo.
\end{acknowledgments}

\appendix
\section{Tensor network representation of Kramers-Wannier duality transformation}
\label{dualTN}

In this appendix, we explain the relationship between the Kramers-Wannier duality \cite{Wu1976DualityTI} and TN representation of the partition function \cite{PhysRevB.81.174411}.
Let us assume that the partition function of the original model can be written down as
\begin{equation}
  Z = \sum_{\{n_i\}} \prod_{\langle i, j\rangle} f(n_i - n_j),
  \label{eq:Z_definition}
\end{equation}
where $n_i=0, 1, \dots, q-1$ and $f(n)$ is a periodic function with a period $q$.
The original lattice forms a directed planer graph and the direction from $i$ to $j$ is specified.
On the other hand, the dual model is defined on the dual lattice.
Its sites are located on each face of the original lattice
and connected by directed edges, the direction of which is a 90° rotation of the original lattice.
The partition function of the dual model is defined as
\begin{equation}
  Z_D = \sum_{\{m_k\}} \prod_{\langle k, l\rangle_D} \tilde{f}(m_k - m_l),
\end{equation}
where $\tilde{f}$ is the Fourier transformation of $f$,
\begin{align}
  \tilde{f}(m) &= \frac{1}{q^{1/2}} \sum_{n=0}^{q-1} e^{-i\frac{2\pi}{q}nm}f(n), \\
  f(n) &= \frac{1}{q^{1/2}} \sum_{m=0}^{q-1} e^{i\frac{2\pi}{q}nm}\tilde{f}(m),
\end{align}
Here, we adopt an uncommon normalization of the Fourier transformation to increase the symmetry of expressions.
The new variable $m_k$ on a site of the dual lattice also takes an integer value from $0$ to $q-1$,
and the summation for $\langle k,l \rangle_D$ runs over all edges of the dual lattice.
In this appendix, we use indices $i$ and $j$ for the original lattice, and $k$ and $l$ for the dual one.
The original (dual) lattice has $N$ ($N_D$) sites and both have $N_B$ edges.
For a finite plane graph, the Euler's relation $N+N_D=N_B+2$ holds.

First, we briefly review the duality transformation based on Ref.~\cite{Wu1976DualityTI} .
Rewriting the partition function in terms of new variables on edges, we have
\begin{equation}
  Z = \frac{1}{q^{N_B/2}} \sum_{\{n_i\}} \prod_{\langle i, j\rangle}
  \left[
  \sum_{m_{ij}=0}^{q-1} e^{i\frac{2\pi}{q}(n_i-n_j)m_{ij}}\tilde{f}(m_{ij})
  \right].
\end{equation}
Summation over $n_i$ yields constraint such that the divergence of $m_{ij}$ at a original lattice site,
\begin{equation}
  \partial_i m \equiv
  \sum_{j\in \partial_i^{\text{(out)}}} m_{ij} - \sum_{j\in \partial_i^{\text{(in)}}} m_{ij},
\end{equation}
is equal to an integer multiple of $q$.
Here, $\partial_i^{\text{(in/out)}}$ is the set of sites connected by edges entering/leaving a site $i$.
Then we can rewrite the partition function as
\begin{equation}
  Z = \frac{q^N}{q^{N_B/2}} \sum_{\{m_{ij}\}} \left(
  \prod_{i=1}^{N} \delta_{\partial_i m, 0}^{(q)} \right)
  \left(
  \prod_{\langle i, j\rangle} \tilde{f}(m_{ij})
  \right).
  \label{eq:Z_tilde_f}
\end{equation}
Now we move to the dual lattice.
Since the edges in the original lattice and the dual lattice are in a common position,
we can regard $m_{ij}$ as a variable $m_{kl}$ on a edge of the dual lattice,
and replace the product over $\langle i,j \rangle$ with the product over $\langle k,l \rangle_D$.
The constraint on $m_{ij}$ becomes that the rotation of $m_{kl}$ around each face on the dual lattice is zero.
It allows us to rewrite $m_{kl}$ using variables $m_k$ on each site of the dual lattice as $m_{kl} = m_k - m_l$.
By removing the overcounting factor $q$ of the mapping between $m_{ij}$ and $m_k$,
we obtain the duality relation,
\begin{equation}
  \frac{Z}{q^{N/2}}
  = \frac{q^{N-1}}{q^{N/2} q^{N_B/2}}
  \sum_{\{m_k\}} \prod_{\langle k,l \rangle_D} \tilde{f}(m_k-m_l)
  = \frac{Z_D}{q^{N_D/2}}.
  \label{eq:duality}
\end{equation}
In the last line, we use the Euler's relation.

The TN representation of $Z$ having the same structure as the original lattice
is usually derived from the eigenvalue decomposition of the local Boltzmann weight on a edge.
Hereafter, we call such TN representation the original representation.
Since the Fourier transformation of $f$ is equivalent to the eigenvalue decomposition,
we can start from Eq.~\eqref{eq:Z_tilde_f}.
By distributing the eigenvalue $\tilde{f}(m_{ij})$ on a edge of the original lattice to connected sites,
we have the original representation as
\begin{align}
  Z &= \frac{q^N}{q^{N_B/2}} \sum_{\{m_{ij}\}}
  \prod_{i=1}^{N} \left[
  \delta_{\partial_i m, 0}^{(q)}
  \left(\prod_{j\in \partial_i} \tilde{f}(m_{ij})^{1/2} \right)
  \right] \\
  &= \frac{q^N}{q^{N_B/2}} \text{tTr} \bigotimes_{i=1}^{N} {\bar{\tau}	}^{[i]},
  \label{eq:Z_TN}
\end{align}
where $\partial_i = \partial_i^{\text{(out)}} \cup \partial_i^{\text{(in)}}$.
The inside of the square brackets gives an element of the local tensor $ {\bar{\tau}}^{[i]}$.
As mentioned in the main text, the representation \eqref{eq:Z_TN} is not appropriate to capture the chiral phase transition.
Thus, we use the different representation of the original model derived in Sec.~\ref{sec2_a}.

We can also obtain another TN representation of $Z$ in the dual lattice, so-called the dual representation.
We introduce a new variable $n_{ij}\equiv n_i - n_j$ as in Sec.~\ref{sec2_b}, 
and replace the summation over $n_i$ in Eq.~\eqref{eq:Z_definition} with one over $n_{ij}$.
This transformation is similar to inverse of the mapping from $m_{kl}$ to $m_{k}$ used in the derivation of the duality transformation.
Thus $n_{ij}$ should satisfy constraint such that its rotation around each face of the original lattice is zero.
In the dual lattice, it imposes the restriction that the divergence of $n_{kl}$ at each dual site is zero.
Therefore, the dual representation of $Z$ is given as
\begin{equation}
  Z = q \sum_{\{n_{kl}\}} \prod_{k=1}^{N_D} \left[
    \delta_{\partial_k n, 0}^{(q)}
    \prod_{l\in \partial_k} f(n_{kl})^{1/2}
  \right],
  \label{eq:Z_TN_dual}
\end{equation}
where the inside of the square brackets gives an element of a local tensor $\tilde{\tau}^{[k]}$ at a dual site $k$.
The factor $q$ comes from the fact that shifting $n_{k}$ simultaneously does not change the value of $n_{kl}$.

Finally, we consider the TN representation of the dual model.
Using the duality transformation \eqref{eq:duality} and the dual representation \eqref{eq:Z_TN_dual}, we have
\begin{equation}
  Z_D = \frac{q^{N_D}}{q^{N_B/2}} \text{tTr} \bigotimes_{k=1}^{N_D} \tilde{\tau}^{[k]},
\end{equation}
where we use the Euler's relation again.
Comparison with Eq.~\eqref{eq:Z_TN} shows that this is nothing but the original representation of $Z_D$.
Since the standard representation of $Z_D$ involves two Fourier transformations,
the local tensor $\tilde{\tau}$ contains the original Boltzmann weight $f$.
Conversely, the dual representation of the dual model is equivalent to the original representation of the original model.

For the AF 6-state clock lattice on the UJ model, the local Boltzmann weight is given as
\begin{equation}
  f(n_i-n_j) = e^{-\beta J \cos\left(2\pi(n_i-n_j)/6\right)}.
\end{equation}
The dual variable $\phi$ defined in the main text corresponds to $2\pi(n_i-n_j)/6$.
Since the dual lattice of the UJ lattice is bipartite, the direction of edges can be set such that all sites of the dual lattice are either all-in or all-out.
Thus we obtain Eq.~(6) as the TN representation of the dual model.

\section{Improved estimation of physical quantity for loop optimization method}
\begin{figure}
  \includegraphics[scale=0.32]{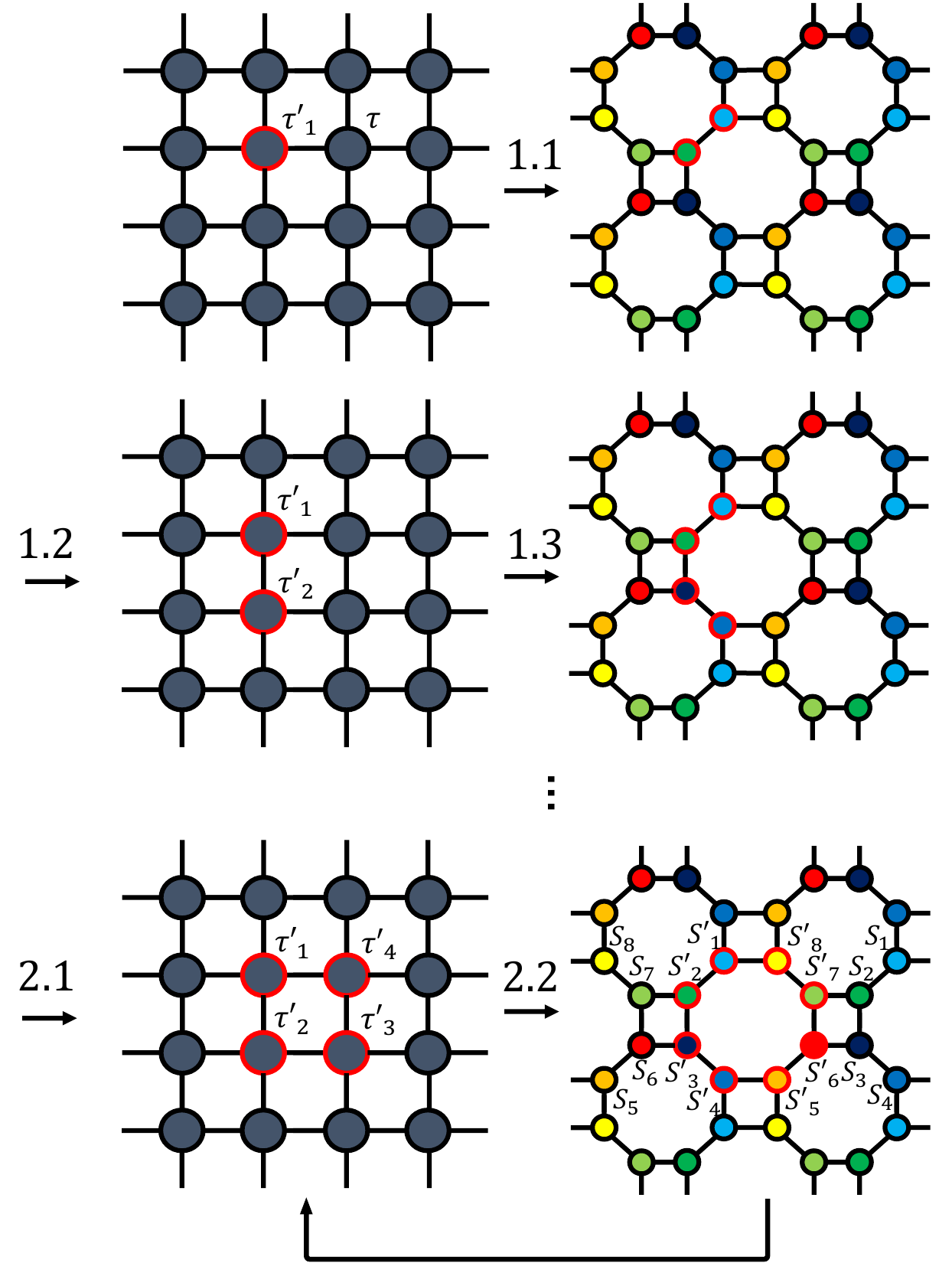}
  \centering
  \caption{\label{fig:impurity}Schematic illustration of loop optimizations on impurity tensors. The gray circles denote the local tensors $\tau$ and the red outlined circles denote the impurity tensors $\tau'$. Throughout the RG steps one can see that the impurity tensors multiply up to $4$ tensors and stop multiplying after the 3rd RG step.} 
  \end{figure}
In this appendix, we will explain some practical methods used in the present NNR-TNR calculation to improve the accuracies of the physical quantities of the single spin valuable and conformal data, which can be used also with other loop-optimization-based methods, such as Loop-TNR \cite{PhysRevLett.118.110504} and TNR+ \cite{PhysRevLett.118.250602}.
\subsection{Loop optimization on impurity tensors}
\label{imp}
To extract the one-point function for a single spin variable at a given site, including chirality, internal energy, etc., it is common to perform the impurity method. In the following, we consider the renormalization scheme of the single impurity tensor using the loop optimization. Fig.~\ref{fig:impurity} shows the renormalization of the single impurity tensor $\tau'$. Up to the first 3rd RG steps, the impurity tensors are multiplied to the square plaquette and later they stop diffusing to the other channel and form a cycle \cite{Nakamoto2016ComputationOC}.
Therefore, it is straightforward to perform the loop optimization on the impurity tensors $S^{'}$ on the octagon configuration, i.e. minimizing $ || \tau'_{1}\tau'_{2}\tau'_{3}\tau'_{4}- S^{'}_{1}S^{'}_{2}S^{'}_{3}S^{'}_{4}S^{'}_{5}S^{'}_{6}S^{'}_{7}S^{'}_{8}||_{F}$, as well as the bulk tensors $S$.

The expectation value of the physical observable $O$ under the periodic boundary can be calculated as,
\begin{eqnarray}
 \label{ape.2}
 \langle O \rangle = \frac{ \mathrm{tTr}(\tau'_{1}\tau'_{2}\tau'_{3}\tau'_{4})}{ \mathrm{tTr}(\tau_{1}\tau_{2}\tau_{3}\tau_{4})}.
\end{eqnarray}

In general, performing the loop optimization on the impurity tensor improves its accuracy. The computational time of this method is simply twice that of the loop-optimization-based TNR algorithms, since the bulk and impurity tensors are optimized twice in a single RG step.

\subsection{Extracting accurate conformal data with moderate computational cost}
\label{lobcg}
Central charge and scaling dimensions of the CFT can be extracted from diagonalization of the linear transfer matrix. When the system can be described by two-dimensional CFT, the partition function of a system size on torus of $L_{x}\times L_{y}$ can be written as
\begin{eqnarray}
 \label{ape.3}
 Z = e^{L_{x}L_{y}f+ O(L_{x}^{-a}L_{y})}\sum_{\alpha}e^{-2\pi\frac{L_{y}}{L_{x}}(\Delta_{\alpha}-c/12)}
\end{eqnarray}
where $f$ is the free energy density, $a$ is some constant larger than one, $c$ and $\Delta_{\alpha}$ are central charge the scaling dimensions. Note that the second exponential term the carries universal information, while the first exponential term has the non-universal contribution $e^{L_{x}L_{y}f}$, which can be set as a normalization factor $N_{f}$ at each RG step.

Increasing the column length of transfer matrix $l_{x}$ should give more accurate conformal data, as the finite-size correction in Eq.(\ref{ape.3}) diminishes. On the other hand, the computational complexity of computing conformal data from $M_{l_{x}}$ in a straightforward fashion (e.g full diagonalization) scales as $O(\chi^{3l_{x}})$. In what follows, we explain how to extract the conformal data accurately, while reducing the computational cost, simultaneously. One strategies is to focus on the largest $N_{\text{eig}}$-th eigenvalues of linear transfer matrix from matrix-vector multiplication. For example, the locally optimal block conjugate gradient (LOBCG) \cite{doi:10.1137/S1064827500366124} is a suitable option as it acquires the largest and leading eigenvalues. In addition, one can further reduce the computational cost by using the 3-leg tensors in octagon configuration after loop optimization. One valid example of the approximation of the linear transfer matrix $M_{4}\ \approx \tilde{M_{4}}$ can be represented as follows
\begin{equation}
\label{ape.5}
\begin{matrix}
 \includegraphics[scale=0.3]{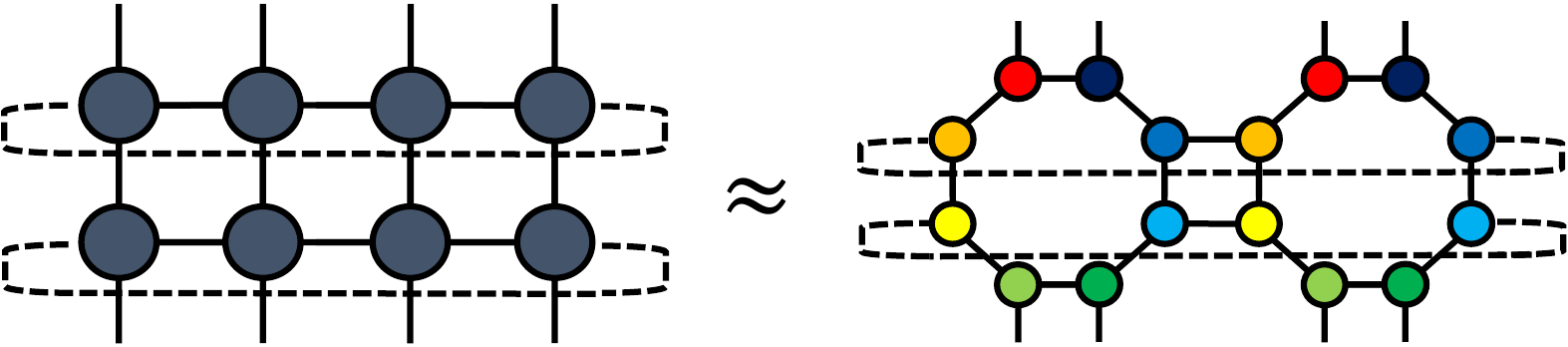}
.\end{matrix}
\end{equation}
Thanks to the loop optimization and the LOBCG method, it allows us to improve the accuracies of conformal data with moderate computational cost, one can solve the eigenvalue problems $\tilde{M}_{4}\vec{x} = \lambda \vec{x}$ using the LOBCG method, or diagrammatically,
\begin{equation}
 \label{ape.6}
 \begin{matrix}
 \includegraphics[scale=0.3]{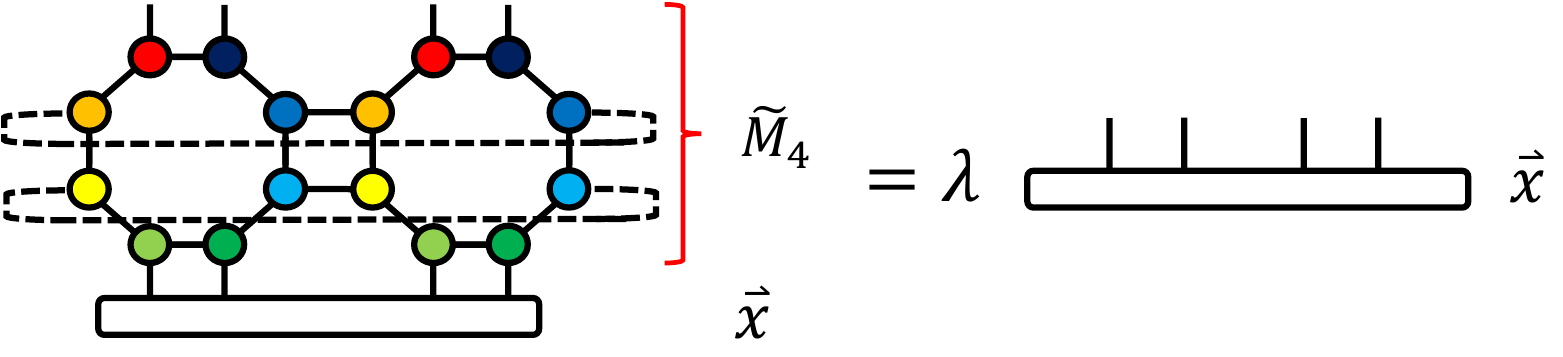}.
 \end{matrix}
\end{equation}
Since the LOBCG method has matrix-free properties, it allows us to compute the eigenvalues and eigenvectors of the approximated linear transfer matrix by effective matrix-vector multiplication $\tilde{M}_{4}\vec{x}$. The overall computational cost of this method is $O(N_{\text{eig}}\chi^{l_{x}+2})$, where $l_{x}$ is an even number.
\bibliography{apssamp}

\end{document}